\documentstyle[twocolumn,prl,aps]{revtex} 
 
\begin{document} 
\twocolumn[\hsize\textwidth\columnwidth\hsize\csname @twocolumnfalse\endcsname
\draft 
\preprint{} 
\title{New universality of metal-insulator transition in
 integer quantum Hall effect system} 
\author{D.N. Sheng, Z. Y. Weng} 
\address{Texas Center for Superconductivity, 
University of Houston, Houston, TX 77204-5506 }  
\maketitle 
\date{tody}
\begin{abstract} 
A new universality of metal-insulator transition in integer quantum Hall
effect (IQHE) system is studied based on a lattice model, where the IQHE states
only exist within a finite range of Fermi energy in the presence of disorders. A two-parameter scaling law is found at the high-energy boundary where direct 
transitions from high IQHE states to insulator occur. We find 
$\rho_{xx}= \rho_{xy}$ at the critical point whose value can continuously
 vary as a single 
function of the Landau-level filling number $n_{\nu}$. Such a new universality 
well explains recent experiment by Song {\it et al.} (Phys. Rev. Lett. ${\bf 78 
}$, 2200 (1997)).
\end{abstract}
]

The global phase diagram proposed by Kivelson, Lee and Zhang\cite{klz} (KLZ) for the quantum Hall effect (QHE) system
has stimulated a series of recent experiments. Although many of these 
experiments are in support of the overall picture of  
KLZ, controversies arise concerning the select rule for 
metal-insulator transitions in such a system\cite{jiang,krv,song}. The global
phase diagram predicts that the transition from an  
IQHE state to insulator can only occur at low filling number 
corresponding to the $\nu=1$ 
IQHE state to insulator ($1\rightarrow 0$) transition, while at higher filling 
number the only allowed transitions are the nearest-neighbor plateau-plateau 
transitions ($\nu \rightarrow \nu\pm 1$).  Experimentally,
however, direct transitions from $\nu=1$, $\nu=2$,  $\nu=3$, and $\nu=6$ IQHE 
states to insulator have all been observed recently\cite{jiang,krv,song} on the 
high-filling number side.
Although the apparent inconsistency of the $\nu=2$ IQHE state to
insulator transition with the global phase diagram  may be reconciled  by  
assuming that the lowest Landau level (LL) is spin degenerate\cite{spi}, the 
higher plateaus to insulator transitions ($3\rightarrow 0$ and $6\rightarrow 0$) 
can not be explained similarly due to the fact that these transitions must take
place at higher LL's.  

The select rule of the global phase diagram is based on
the so-called float-up picture\cite{khm,lau} where extended states in the weak
 field limit are 
assumed to shift towards higher energy  indefinitely without disruption. But 
recent numerical calculations\cite{lui,dn} have shown that the float-up picture is not 
correct in a lattice model in which extended states merge and disappear
before they can reach to the band center. This non-float-up picture 
holds down to very weak magnetic field limit\cite{dn} which means that the 
lattice model is not equivalent to the continuum model even in the limit where 
the magnetic length scale is much larger than the lattice constant. This 
surprising result, which can be well understood based on a topological Chern 
number description\cite{dn}, implies that the lattice effect will always remain 
to be an 
important factor in the problem of metal-insulator transitions even in the weak 
field limit. It may thus provide an explanation for the aforementioned 
experimental observations of a direct transition from the higher IQHE states to 
insulator.

Recent experimental measurement has further indicated\cite{song}  that 
the direct transition from a higher IQHE state to insulator may belong to a new 
universality. In the corresponding critical regime, the longitudinal and transverse conductances
generally satisfy $\sigma _{xxc}=\sigma_{xyc}$ within the experimental error 
and their value varies continuously with the filling number, which is different
from the usual plateau-plateau transitions (including $1\rightarrow 0$
transition as well) in which  $\sigma_{xxc}=0.5 e^2/h$ and $\sigma_{xyc}=(\nu+ 
0.5) e^2/h$\cite{con}. So far such a new class of metal-insulator transitions 
in the IQHE system cannot be understood by any existing theory of the IQHE based 
on the continuum model. On the other hand, since the direct transitions from 
higher IQHE states to insulator are present naturally in a lattice model as 
mentioned above, it would be very desirable to examine the corresponding 
critical behaviors. 

In this paper, we present a numerical study 
based on the calculation of longitudinal and transverse conductances and a finite-size scaling analysis near these 
critical points. Our results show that a direct metal-insulator transition
in this regime indeed belongs to a new universality. In particular, the 
scaling behavior of localization length near each critical point obeys a real
{\it two-parameter} scaling law, in contrast to the plateau-plateau transitions 
whose critical behavior can be reduced to a one-parameter scaling law\cite
{huk}. As the
consequence, $\sigma_{xxc}$ and $\sigma_{xyc}$, which are equal 
to each other,  continuously 
vary as a function of a single 
parameter --- the LL filling factor $n_{\nu}$, and does not depend on magnetic 
field or disorder strength independently. Since such a critical behavior is 
independent of the strength of magnetic field, we are able to make a close
comparison of our numerical results with experimental measurement\cite{song}
and find a consistent explanation. 

The lattice Hamiltonian to be studied is a tight-binding model (TBM) given as
follows: 
\begin{eqnarray*} 
H=-\sum_{<ij> } e^{i a_{ij}}c_i^+c_j + H.c. +\sum _i w_i c^+_i c_i , \nonumber
\end{eqnarray*} 
where the hopping integral  is taken as the unit, and $c_i^+$ is a  fermionic creation 
operator with $<ij>$  referring to two  nearest neighboring sites.  A uniform  
magnetic flux per plaquette is  given as  $\phi=\sum _ {\Box} a_{ij}=2\pi/M$, where the  
summation runs over four links around a plaquette.
$w_i$ is  a random potential with strength $|w_i|\leq W/2$,  
 and the
white  noise limit is considered with no correlations among different  sites for $w_i$. 

In the weak disorder limit, a well-defined IQHE 
plateau structure is exhibited in the Hall conductance.
For our purpose, the disorder strength $W$ is continuously increased to 
identify a direct transition from the $\nu$th IQHE state to 
insulator. For instance, in Fig. 1(a), by fixing the Fermi energy near the center 
of the $\nu=2$ plateau ($E_f=-2.75$) at weak disorder, we show how the Hall 
($\sigma_{xy}$) and longitudinal ($\sigma_{xx}$) conductances evolve with 
disorders (at a flux strength $\phi=2\pi/16 $). Here $\sigma_{xy}$ is calculated based on the Kubo formula and $\sigma_{xx}$ by the Landauer formula\cite{lan}.   
At weak disorder ($W\le 1$), $\sigma_{xy}$ is well-quantized at $\nu e^2/h$
with $\nu=2$. With $W$  increasing from 1 to 8, $\sigma_{xy}$ continuously
reduces to zero without showing a $\nu=1$ plateau, indicating a direct
$\nu=2\rightarrow \nu=0$ transition. Two curves for $\sigma_{xy}$ at  different
sample sizes ($N=16^2$ and $N=32^2$) cross at a critical point which also
coincides with the maximum point of the longitudinal conductance $\sigma_{xx}$.
The latter is also sample-size independent at the corresponding critical
disorder strength $W_c=3.5$. But away from $W_c$, $\sigma_{xx}$ monotonically 
decrease with the increase of sample size as  a typical behavior in localized states. Thus the existence of a single $W_c$ identified from both $\sigma_{xy}$ and $\sigma_{xx}$ confirms that the $\nu=2\rightarrow 0$ transition is really a
one-step transition. Furthermore, a calculation of the finite-size localization
length by using the transfer matrix method\cite{mac}
 also verifies the same $W_c$:  
at the fixed $E_f=-2.75$, we find that the localization length satisfying the 
scaling relation, $\lambda_L/L=f'(\xi(W)/L)$, for a stripe sample with the width varying 
from $L=16,24, 32, 48, 64$, to $80$. Here $\xi$ is identified  as the
thermodynamic localization length which diverges at $W_c=3.5$ as shown in 
Fig. 1(b). At such a critical point, we obtain  $\sigma_{xxc}=(1.06\pm 0.02) 
e^2/h $ and $\sigma_{xyc}=(1.01 \pm 0.06) e^2/h$, or,
$\sigma_{xxc}=\sigma_{xyc}$ within the statistics error bar ($6\%$).

Fixing the Fermi energy $E_f$ near the center of the $\nu$th plateaus at weak 
$W$, we always find a similar direct transition to insulator ($\nu\rightarrow 
0$) as $W$ increases. The critical disorder $W_{c\nu}$ for $\nu \rightarrow 0$
transition is found to satisfy a sequence $W_{c\nu}<W_{c\nu'}$ for $\nu>\nu'$. 
For instance, we have $W_{c1}=3.8$, $W_{c2}=3.5$, $W_{c3}=3.2$ and $W_{c4}=2.8$
at a flux strength $2\pi/16$. Up to the weakest flux strength $2\pi/384$ that we 
can get access to, we always find the same sequence for the disappearance of the 
IQHE plateaus. According to this sequence, near the critical disorder $W_{c\nu}$ 
for $\nu$th plateau, all higher plateaus have already disappeared while lower ones are still 
robust. We can then define the transition of $\nu\rightarrow 0$ as the 
boundary of the IQHE regime on the high-energy (or high-filling number) side, 
beyond which the IQHE states no longer exist up to the band center. The 
corresponding critical conductances $\sigma_{xxc}$ and 
$\sigma_{xyc}$ at $W_{c\nu}$ always satisfy the relation $\sigma_{xxc}=\sigma_{xyc}$ within the  numerical error bar, and their values continuously vary with the plateau index $\nu$ and the Fermi energy $E_f$.
Nevertheless, we find that all of them can be scaled into a single 
curve as the function of the LL filling number $n_{\nu}$ shown in Fig. 2,
in which the magnetic flux changes from $2\pi/16$ to $2\pi/96$.    
In Fig. 2, we show $\rho _{xxc}$ and $\rho _{xyc}$ instead of $\sigma_{xxc}$ and 
$\sigma_{xyc}$ in order to compare with experimental data of Song, {\it et 
al.}\cite{song} in the insert of Fig. 2. 
We see that the overall agreement between the theory 
and experiment is quite reasonable. (Note that in the insert the
top horizontal axis is carrier density which has not been converted to  
$n_{\nu}$ here as the corresponding magnetic fields are not provided in
Ref. \onlinecite{song}.)    
Both  $\rho _{xxc}$ and $\rho_{xyc}$ monotolically increase with the decrease of
$n_{\nu}$ and are eventually saturated at $h/e^2$. 
We can make a closer comparison:  in Ref. \onlinecite{song}  the $\nu=3 
\rightarrow 0$
transition covers a range of critical resistivities from $0.27$ to $0.4  $ in 
unit of $h/e^2$, which agrees well with the theoretical values shown in Fig. 2.  
The theory also predicts that the transitions reported in the experiment  with
critical  resistivities less than $0.27$ ($h/e^2$) actually come from higher 
($\nu \ge 4$) IQHE states to insulator transitions. 

Thus we have identified a new universality of metal-insulator transition at 
the boundary between the IQHE regime and insulator on the
{\it high} filling-number (energy) side in the TBM. It is quite different from 
the well-known  
plateau-plateau transition ($\nu\rightarrow \nu\pm 1$) within the IQHE regime
as well as the transition from the IQHE state to insulator on the low 
filling-number side (i.e., $\nu=1\rightarrow 0$)
where, at $\nu\rightarrow \nu+1$ transition, one has $\sigma_{xxc}\simeq 0.5 
e^2/h$, while $\sigma_{xyc}=(0.5+\nu) e^2/h$\cite{con}. It provides a unique
explanation for recent experiments\cite{song,krv}. In order to further 
understand this new universality of critical behavior at the high-filling 
boundary of the IQHE, we investigate the scaling behavior in this regime below.

According to the general scaling theory of the QHE system\cite{pru}, the 
finite-size localization length may be written as a general
function of two parameters, i.e., $\lambda_L/L=f(L/\xi, p)$, at a large 
sample with a width $L$. Here the 
first parameter is $L$ divided by the thermodynamic localization 
length $\xi$ as 
usual\cite{mac}, and the second one, $p$, may be chosen as either $\sigma_{xx}$ 
or $\sigma_{xy}$  at a fixed sample size $L_0$.  In the present case, it is 
convenient to choose $\sigma_{xxc}$ as $p$ (which can be connected with 
$\sigma _{xx}$ and $\sigma_{xy}$ at $L=L_0$ by the scaling-flow 
diagram{\cite{pru}). 
Based on this scaling hypothesis, the numerical data at
different flux strength $2\pi/M$'s, disorder strength $W$'s, sample size $L$'s,
etc.,  should all collapse into a single curve so long as the critical conductance $\sigma_{xxc}$ is the same.
Indeed, as shown in Fig. 3, the data of finite-size localization lengths can be 
well fit into one curve by choosing a single scaling variable $L/\xi$ at the
given $\sigma_{xxc}=1.06$. Such a $\sigma_{xxc}$ corresponds to a 
$\nu=2\rightarrow 0$ transition. By changing the plateau index $\nu$ or the 
Fermi energy $E_f$, the critical conductance $\sigma_{xxc}$ can change
continuously as discussed before. One expects the scaling curve to change
correspondingly. In the insert of Fig. 3, scaling curves at two critical points 
($\nu=2\rightarrow 0$ and $\nu=4\rightarrow 0$) are shown with 
$\sigma_{xx2}=1.06$ and $\sigma_{xx4}=2.1$, respectively. Different scaling curves have also been obtained for the transitions of $\nu=3\rightarrow
0$ and $\nu=1\rightarrow 0$ in this
regime. All of these different cases can be generally specified by a single 
LL filling number $n_{\nu}$, since $\sigma_{xxc}$ is uniquely determined 
by $n_{\nu}$ as shown in Fig. 2. Therefore, a two-parameter scaling law is 
well established here which
characterizes a new universality of metal-insulator transitions between
the IQHE regime and insulator at high-filling-number boundary. It is noted that 
the  transition from the IQHE state to insulator at low-filling-number boundary 
(corresponding to $\nu=1\rightarrow 0$) is usually well described by a 
one-parameter scaling law\cite{huk}, which may be understood as the second 
parameter $\sigma_{xxc}$ always remains a constant in this regime\cite{con}. So 
the present case is the first one in the  QHE systems that a two-parameter 
scaling law becomes 
necessary in order to describe the critical behavior. Finally, if one plots the 
finite-size conductance $\sigma_{xxL}$ as a function of the finite-size 
localization  length $\lambda_L/L$ at different flux strengths ($2\pi/16- 
2\pi/96$), as well as different plateau index 
$\nu$'s, the Fermi energy $E_f$'s and disorder strength $W$'s,  all the data 
also collapse into one curve as shown in Fig. 4. (The notations of the data in Fig. 4 are similar to those in Fig. 3.) It means that $\sigma_{xxL}$ 
is uniquely decided by $\lambda_L/L$,  and thus also satisfies the two-parameter 
scaling law.

Finally, we would like to point out that in the above discussion, we have not 
considered the spin degree of freedom. For the conventional plateau-plateau 
transition, it is well-known\cite{lee} that the mixing of LL's for different 
spins by a spin-orbit coupling will not change the universality class and such a 
spin-coupled system behaves still like a {\it spinless} system. In our case, we have checked that introducing a weak 
spin-orbit coupling effect will also make the system with spin degree of freedom 
behave like a spinless system. For example, the $\nu=2 
\rightarrow 0$ and $\nu=3\rightarrow 0$ transitions remain to be direct transitions (even with Zeeman splitting) with the same universality of critical 
behavior discussed above for the spinless case. Both diagonal and Hall 
conductance are also equal to each  other and fall into the same range shown in 
Fig. 2. Details will be presented elsewhere.

In summary, we have identified a new two-parameter scaling law in the 
critical regime of direct transitions from the high IQHE states to insulator in 
the TBM. Such a new universality of metal-insulator transitions predicts that
$\rho_{xxc}=\rho_{xyc}$ and their value depends solely on the LL filling number 
$n_{\nu}$. The critical behavior of this lattice model in the aforementioned direct
phase transition regime provides a consistent explanation for recent experimental measurements. 

{\bf Acknowledgments} -The authors would like to thank T. Xiang, D. Shahar, 
D. Kravchenko, K. Yang, 
and C. S. Ting for stimulating and helpful discussions.
The present work is supported by Texas Center for Superconductivity at University of Houston, and a grant from Robert Welch foundation.

Fig. 1 (a) The evolution of $\sigma _{xx}$ and $\sigma_{xy}$ with disorder strength $W$ at a fixed Fermi energy $E_f=-2.75$, which corresponds to the 
center of the $\nu=2$ IQHE plateau at weak $W$ limit. Lattice sizes vary form 
$16\times 16$ ($* $), $32\times 32$ ($+$), to  $64\times 64$ ($\bullet $), and 
the sample-independent fixed point at $W_c=3.5$ indicates the crossing of the 
last extended level at $E_f$. (b) Thermodynamic localization length  $\xi$ also 
diverges at $W_c$.  

Fig. 2 Critical resistivities, $\rho_{xxc}$ and $\rho_{xyc}$,
as function of Landau-level filling number $n_{\nu}$, respectively, at flux 
strength $\frac{2\pi}{16}$ ($\triangle$, $+$); $\frac{2\pi}{24}$ ($\diamond$, $\times$); $\frac{2\pi}{32}$ ($\Box$, $*$) and $\frac{2\pi}{96}$ ($\bullet$, 
none). The insert: the experimental data by Song, {\it et al.}\onlinecite{song}.
 
Fig.3 Two-parameter Scaling function $\lambda_L(L/\xi, \sigma_{xxc})/L$  versus  $L/\xi$  at $\sigma _{xxc}=1.06$. All the data at
different disorders collapse into one curve with the stripe sample 
width $L$ varying from $L=16$ to $80$ and flux strengths from $\frac{2\pi}{16}$ 
($*$), $\frac{2\pi}{24}$ ($\diamond $), $\frac{2\pi}{32}$ ($\Box$) to $\frac{2\pi}{96}$ ($\triangle$).
The insert: two different scaling curves for the $\nu=2 \rightarrow 0$
(at $\sigma_{xx2}=1.06$) and $\nu=4 \rightarrow 0$ (at $\sigma_{xx4}=2.10$)
transitions, respectively. 

Fig. 4. Finite-size conductance $\sigma_{xxL}$ as a single function of 
$\lambda_L/L$ at different sample-sizes, flux strengths, plateaus indexes,
disorder strengths, and Fermi energies.

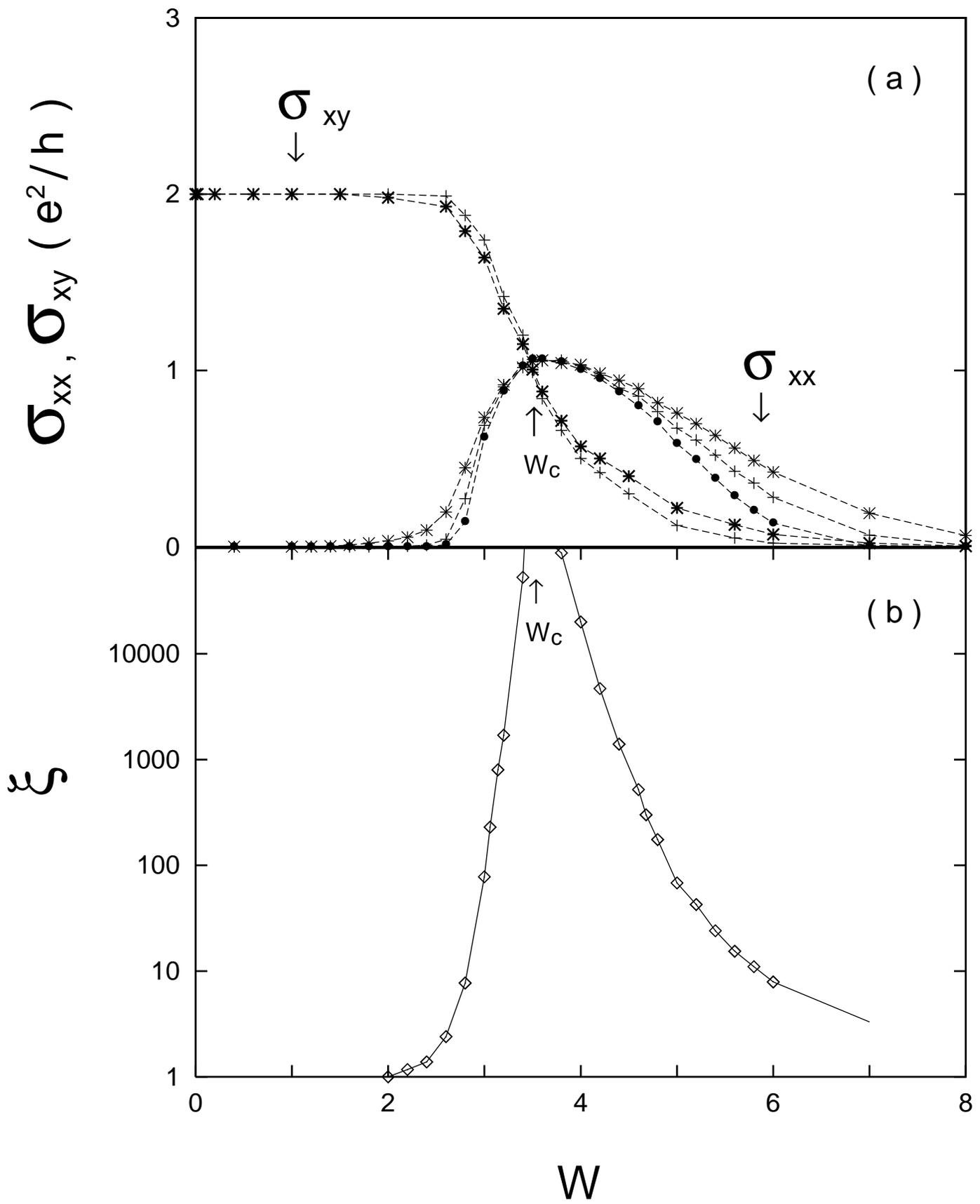
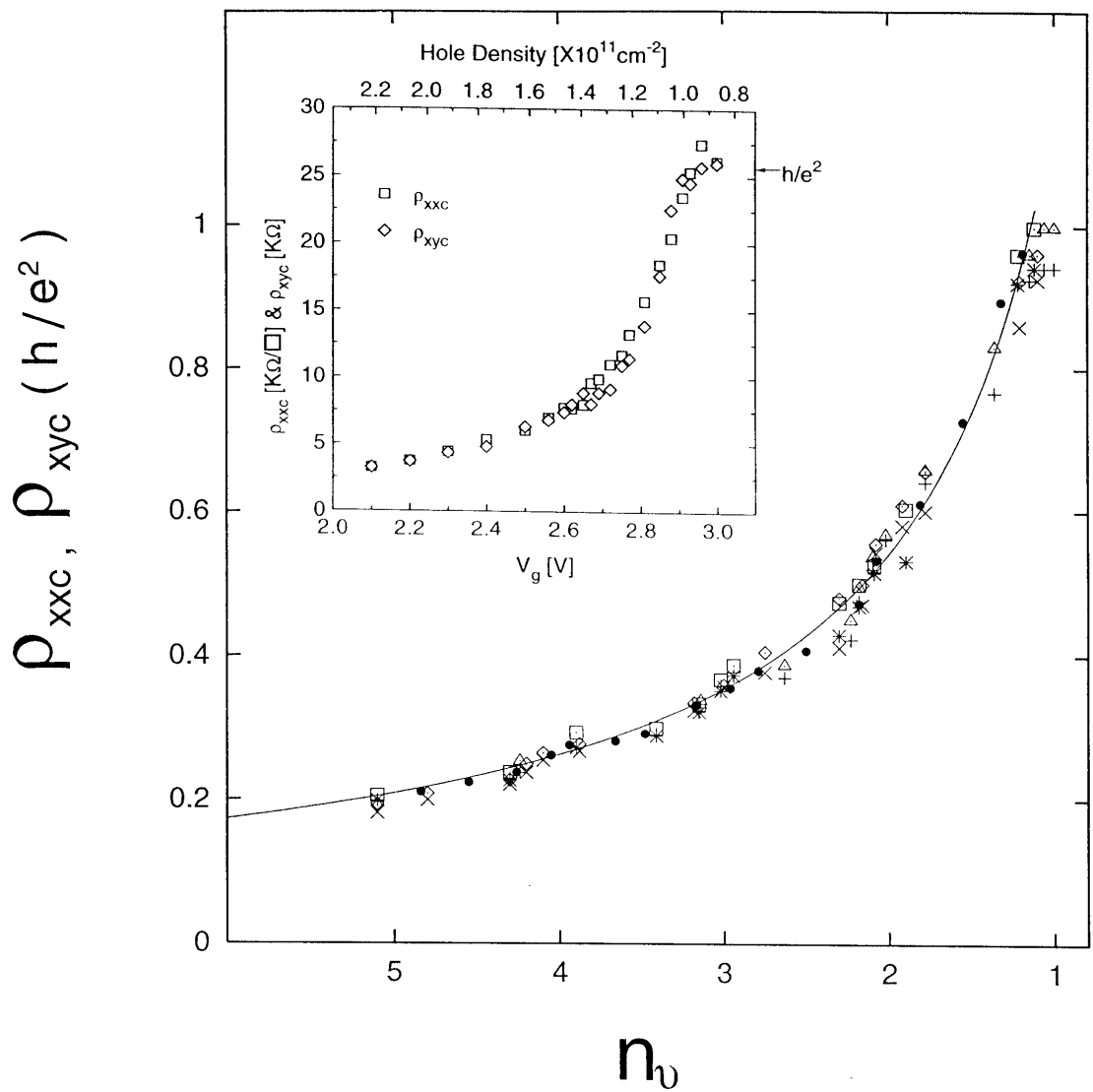
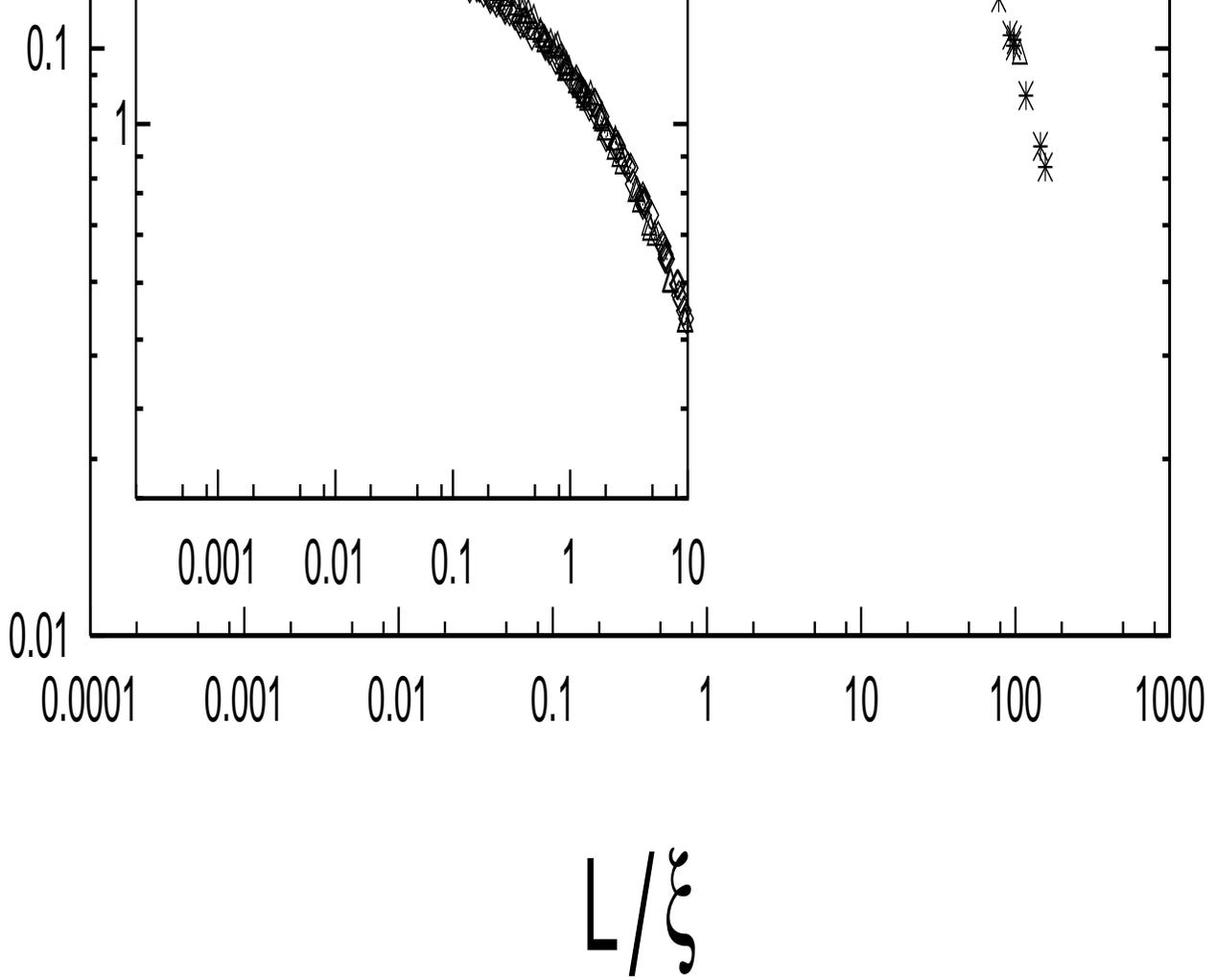
\end{document}